\documentstyle{pasj00}
\begin{document}

\SetRunningHead{A.Imada et al.}{A new WZ Sge star, TSS J022216.4+412259.9}

\title{Discovery of a new dwarf nova, TSS J022216.4+412259.9: \\
 WZ Sge-type dwarf novae \\
breaking the shortest superhump period record}

\author{Akira \textsc{Imada},$^1$Kaori \textsc{Kubota},$^1$Taichi
\textsc{Kato},$^1$Daisaku \textsc{Nogami},$^2$Hiroyuki
\textsc{Maehara},$^3$ \\
Kazuhiro \textsc{Nakajima},$^4$Makoto \textsc{Uemura},$^5$ and Ryoko
\textsc{Ishioka}$^6$}

\affil{$^1$Department of Astronomy,Faculty of Science, Kyoto University,
       Sakyo-ku, Kyoto 606-8502}
\affil{$^2$Hida Observatory, Kyoto University, Kamitakara, Gifu 506-1314}
\affil{$^3$VSOLJ, Namiki 1-13-4, Kawaguchi, Saitama 332-0034, Japan}
\affil{$^4$VSOLJ, 124 Isatotyo Teradani, Kumano, Mie, Japan}
\affil{$^5$Hiroshima Astrophysical Science Center, Hiroshima University,
       Hiroshima 739-8526, Japan}
\affil{$^6$Subaru Telescope, National Astronomical Observatory of Japan
       650 North A'ohoku Place, Hilo, \\ HI 96720, U.S.A.}
\email{a\_imada@kusastro.kyoto-u.ac.jp}

\KeyWords{
          accretion, accretion disks
          --- stars: dwarf novae
          --- stars: individual (TSS J022216.4+412259.9)
          --- stars: novae, cataclysmic variables
          --- stars: oscillations
}

\maketitle

\begin{abstract}
We report on the time-resolved CCD photometry of a newly discovered
 variable star, TSS J022216.4+412259.9 during the outburst in 2005
 November-December brightening. The obtained light curves unambiguously
 showed 0.2-0.3 mag modulations, which we confirmed to be the superhump
 observed among SU UMa-type dwarf novae. We also performed  a period
 search for the data obtained during the outburst plateau phase, and
 revealed the existence of the two periodicities: 0.054868(98) days for
 the first two nights and 0.055544(26) days for the following plateau
 phase. This bi-periodicity is hardly observed in usual SU UMa-type
 dwarf novae, but characteristic of WZ Sge-type stars. We undoubtedly
 detected a rebrightening in the post-outburst stage, which is typical
 of short-period SU UMa-type dwarf novae including WZ Sge-type
 stars. These observations suggests that TSS J022216.4+412259.9 may be a
 new WZ Sge stars breaking the shortest superhump period of 0.05648 days
 for V592 Her among this class with a known superhump period so far.
\end{abstract}

\section{Introduction}

Since the advent of the high speed CCD photometry, many studies have
been performed concerning the variability of dwarf novae
(\cite{war95book}; \cite{kat04vsnet}), which have led a significant
progress in understanding many aspects of dwarf novae-outburst.
WZ Sge stars, an extreme subclass of SU
UMa-type dwarf novae, show many peculiarities both in outburst and
quiescence. One of the enigmatic phenomenon is that they exhibit so
called ``early superhumps'' on the early stage of the superoutburst of
WZ Sge stars (\cite{kat96alcom}; \cite{osa02wzsgehump}; \cite{kat02wzsgeESH};
\cite{pat02wzsge}). The period of the humps is almost identical to that
of the orbital period of the system. Another peculiarity is that most of
WZ Sge stars show rebrightenings (\cite{kuu96TOAD}; \cite{pat98egcnc})
after the main outburst. Extensive monitoring of WZ Sge stars, as well
as insightful theoretical works, have shed some light on elucidating
these origin and revealing the diversity of them, while less than
ten objects have been certified as this class. However, interpretation
of the aforementioned phenomena is controversial.

Recently, there has arisen a new problem regarding the evolution of
dwarf novae. According to a population synthesis, there should
be more dwarf novae below the period gap than observed
\citep{how97periodminimum}. \citet{how97periodminimum} further suggested
that almost all of CVs should have the orbital period below the period
gap. However, observations indicate that the number of CVs above the period
gap is comparable to that of CVs below the period gap
\citep{gan05review}. Another calculation suggests the ``theoretical
period minimum`` of dwarf novae will lie as short as 64 min
\citep{kol99CVperiodminimum} under the secular
evolution. \citet{wil05CVevolution}, on the other hand, has shown that
the ``period minimum'' becomes around 80 min by an inclusion of an
effect of a circumbinary disk. After reaching the ``period minimum'',
the system comprises a degenerate secondary, which gives rise to
inversion of the mass-radius relation. Then the orbital period of the
system will become longer as the CV evolves. Meanwhile, well accepted is
that the ``observational period minimum'' lies around 78
min.\footnote{V485 Cen and EI Psc, recently discovered dwarf novae with
the periods breaking the minimum, are suggested to have experienced
another evolution with an evolved secondary
\citep{pod03amcvn}.}\footnote{Theoretical works suggest the value of the
period minimum depends on chemical abundance of the secondary star.} As far
as the evolutional scenario of dwarf novae is concerned, intensive
studies should be awaited especially from the theoretical side (see
\citet{how01periodgap} for a review.).

TSS J022216.4+412259.9 was firstly discovered by \citet{qui05atel658} as
a candidate of an ongoing supernova on 2005 November 16.1 (UT) with about
15.5 mag. The object maintained almost the same magnitude until 2005
November 17.1 (UT), which was confirmed by 45 cm ROTSE-IIIb telescope at
the McDonald Observatory. Negative observations were carried out at the
same site, giving 16.5 mag as an upper limit on 2005 November 15.1
(UT). On 2005 November 18, an optical spectrum was obtained with a 2.2 m
telescope located on the University of Hawaii, which revealed the dwarf
nova nature of the eruptive object by unambiguous detection of
H$\alpha$ and HeII 4686~\AA in emission, as well as a blue
continuum \citep{qui05atel658}. There is no optical counterpart in USNO
B1.0 and 2MASS catalog within the 5 arcsec circle. This suggests that
the quiescent magnitude of the object is fainter than 21 mag.

In this letter, we report on the results of CCD photometric observations
during the 2005 November long outburst of TSS 022216.4+412259.9
(hereafter, aliased TSS J0222), mainly focusing on detection of two
periodicities during the outburst plateau stage. Detailed discussion
will be published in the forthcoming paper.

\section{Observations}

\begin{figure}
\begin{center}
\resizebox{80mm}{!}{\includegraphics{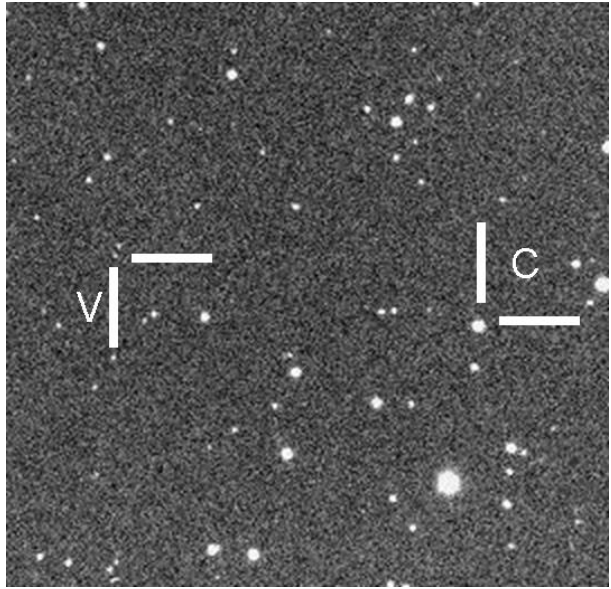}}
\end{center}
\caption{Finding chart of TSS J0222 from our original data. The field of
 view is about 16'${\times}$8'. The north is up and the east is to the
 left. The variable and the comparison star are denoted by $V$ and $C$,
 respectively.}
\end{figure}

\begin{figure}
\begin{center}
\resizebox{80mm}{!}{\includegraphics{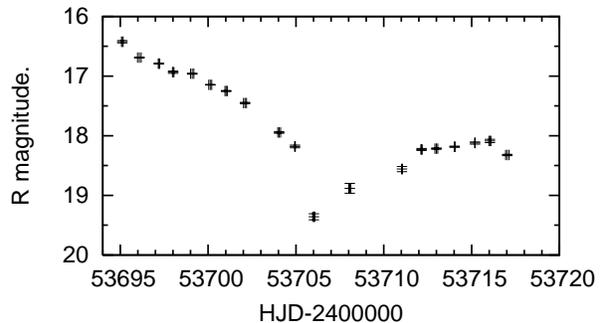}}
\end{center}
\caption{The daily averaged light curve during the outburst. The vertical
 and horizontal axes denote $R$ magnitude and HJD,
 respectively. The magnitude of the comparison is $R$ = 12.6. The
 error bar denotes standard error. Due to the faintness of TSS J0222,
 the accuracy of the magnitude calibration was dependent on the
 weather condition so that we excluded a few data. Nevertheless, we
 unambiguously detected the rebrightenings around HJD 2453708.}
\label{}
\end{figure}

Time-resolved CCD photometry was carried out from 2005 November 21 to
2005 December 14 at 4 sites using 25-60 cm telescopes. More information
about each site is described in table 1. The total data
points exceeded 10000. Typical exposure time was 10-30 sec with no
filter. The resultant band pass is close to R$_{\rm c}$-band. The differential
magnitude among each sites was adjusted to that of the Kyoto site. We
used USNO$-$A2.0 1275$-$01425512 ($B$ = 12.8, $R$ = 12.6) as a
comparison star, whose constancy was checked by several stars in the
same image. Figure 1 illustrates a finding chart of TSS J0222. Kyoto
team used java-based PSF photometry package developed
by one of authors (TK). Hida and Saitama used IRAF DAOPHOT and
APPHOT. Data of Mie were analysed by
FitsPhot4.1.\footnote{http://www.geocities.jp/nagai\_kazuo/dload-1.html}
Because of the faintness of TSS J0222, accurate calibration of the
magnitude was prevented on some nights. However, such uncertainty would
not influence on the following analysis in which we mainly focus on the
periodicity of photometric variations. Heliocentric correction is also
made for our whole run.

\begin{table}[htb]
\caption{Information about each site.}
\begin{center}
\begin{tabular}{cccc}
\hline\hline
Site & Telescope & Exp(s)$^\dagger$ & N(days) \\ 
\hline
\hline
Kyoto, Japan & 40cm & 30 & 20 \\
Hida, Japan & 60cm & 10 & 1 \\
Saitama, Japan & 25cm & 30 & 2 \\
Mie, Japan & 25cm & 30 & 1 \\
\hline
\multicolumn{4}{l}{$^\dagger$ Exposure time.} \\
\multicolumn{4}{l}{$^\ddagger$ Number of observed nights.} \\
\end{tabular}
\end{center}
\end{table}

\section{Results and discussion}

\subsection{light curves}

Figure 2 displays the overall light curve of this outburst. At the onset
of our observations, TSS J0222 was at the magnitude of 16.4. After that
the magnitude linearly faded down with the rate of 0.1
mag d$^{-1}$. Some WZ Sge stars show a rapid decline just after the
maximum brightness with $\sim$ 0.5 mag d$^{-1}$, presumably due to
viscous decay in the accretion disk \citep{kat02v592her}. In the case of
TSS J0222, we cannot specify such a rapid declining took place due to
the lack of the our run at the early stage of the superoutburst. After
the end of the main plateau stage, TSS J0222 faded down as faint as our
detection limit ($\sim$ 19.5 mag) on HJD 2453706. However, we
detected a rebrightening on HJD 2453708 at the magnitude of
18.8. Judging from figure 2, the rebrightening lasted at least 9
days. Because of an intrinsic faintness of TSS J0222, we could not
specify the overall feature of the rebrightenings. Phenomenologically,
three types of rebrightenings may exist. AL Com and WZ Sge itself kept
bright for a week or even more, although about 1 mag-fluctuations were
observed with a time scale of a few days in WZ Sge. The overall light
curves of rebrightenings are not discrete, but somewhat
long-lasting. Here we call this type of rebrighenings ``type-A''. On the
other hand, regular rebrightenings were seen in EG Cnc. Here we call
this kind of rebrightenings ``type-B''. Finally, not only WZ Sge stars
but also some SU UMa stars show a short rebrightening. Such systems
include RZ Leo \citep{ish01rzleo}. We call ``type-C'' for the
rebrightening. The observed rebrightening of TSS J0222 apparently
belongs to the "type-A" category. The reason why different
rebrightenings emerge even in the same system, is not known (for a
relevant paper, see e.g., \cite{kuu96TOAD};
\cite{osa01egcnc}).footnote{For example, WZ Sge shows two types of
rebrightenings. In 1978 and 2001 superoutburst, the object exhibited the
type-A rebrightening, while no rebrightening was observed in 1946
superoutbutst (\cite{pat81wzsge}; \cite{ish02wzsgeletter}}.

\begin{figure}
\begin{center}
\resizebox{80mm}{!}{\includegraphics{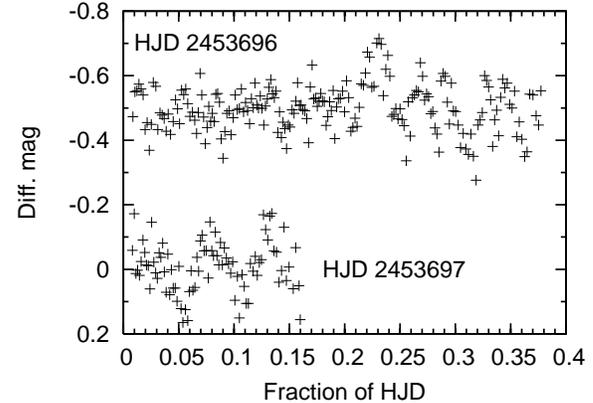}}
\end{center}
\caption{Light curves on HJD 2453696 and HJD 2453697, second and third
 night of our observations. Both light curves were subtracted a linear
 declining trend. Note that there are vague features on HJD 2453696,
 while there are features of superhumps on HJD 2453697.}
\label{}
\end{figure}

Figure 3 displays enlarged light curves on HJD
2453696 and HJD 2453697. There is almost no feature on HJD
2453696 at a glance. However, double-peaked features can be seen when
the data of HJD 2453695 and HJD 2453696 are folded with 
0.054868 days (discussed later). As for the light curve on HJD 2453697,
we can clearly see a rapid-rise and a slow decline feature, which is
typical for the superhump profile. We suggest that some kind of
transition took place between HJD 2453696 and HJD 2453697, and that the
feature we observed on HJD 2453697 was genuine superhumps, as is
observed in SU UMa stars during superoutbursts. This kind of
transition has been observed in WZ Sge stars amid the plateau
stage (e.g., \cite{nog97alcom}; \cite{pat98egcnc};
\cite{ish02wzsgeletter}). Therefore, we can reasonably conclude that we
did observe early superhumps on HJD 2453695 and HJD 2453696, and genuine
superhumps from HJD 2453697 on, respectively.

\subsection{period analysis}

\begin{figure}
\begin{center}
\resizebox{80mm}{!}{\includegraphics{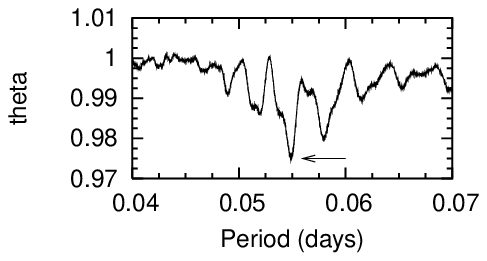}}
\resizebox{80mm}{!}{\includegraphics{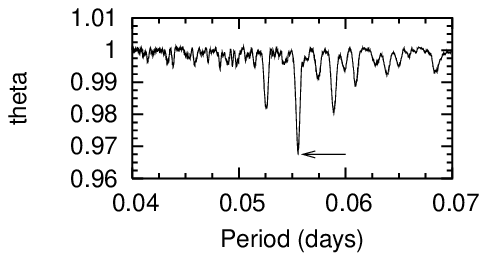}}
\end{center}
\caption{Theta diagrams of the period analyses using the PDM method for
 the data obtained during the first two days (Top panel) and the
 following nights of the plateau stage (Bottom panel). For the first two
 days, 0.054868(98) days is the best estimated period, while the
 periodicity of 0.055544(26) days is shown since HJD 2453697. We
 interpreted the latter as the mean superhump period of TSS J0222.}
\label{}
\end{figure}

\begin{figure}
\begin{center}
\resizebox{80mm}{!}{\includegraphics{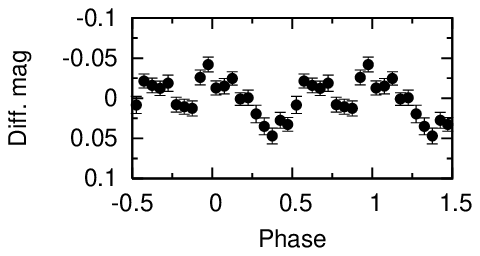}}
\resizebox{80mm}{!}{\includegraphics{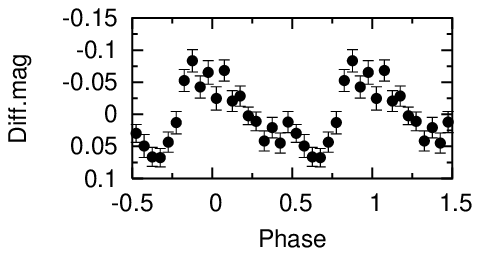}}
\end{center}
\caption{Phase-averaged light curves during the two stages folded with
 the above mentioned period, 0.054868 days and 0.055544 days,
 respectively. One can see double-peaked profiles, as is observed in WZ
 Sge-type dwarf novae in the top panel. On the other hand, a rapid
 rise and slow decline, characteristic of superhumps, is evident on the
 bottom panel.}
\label{}
\end{figure}

The results of period analyses during the plateau phase are represented in
figure 4. After subtracting a linear decline trend for each run, we performed
the phase dispersion minimization method (PDM, \citet{ste78pdm}) for the
data obtained on HJD 2453695 and HJD 2453696, corresponding to the early
superhump phase (upper panel), and for following 7 days, corresponding
to the genuine superhump phase (lower panel), respectively. The
statistical errors were estimated by using Lafler-Kinman class method
\citep{fer89error}. As can be obviously seen in figure 4, two different
periodicities were exhibited: 0.054868(98) days for early superhumps,
and 0.055544(26) days for genuine superhumps, respectively. The
statistical errors were estimated by using Lafler-Kinman class method
\citep{fer89error}. The statistical F-tests yielded significance levels
of the signals 74$\%$ for early superhumps and 83$\%$ for genuine
superhumps, respectively. Note that
the resultant superhump period is the shortest record among WZ Sge
stars, exceeding 0.05648(2) days of V592 Her
\citep{kat02v592her}. Figure 5 shows averaged light curves during these
stages folded by the periods obtained above. Note that the profile of
the upper panel is not singly peaked, but definitely doubly peaked,
similar to that observed in the early phase of WZ Sge and AL Com
\citep{ish02wzsgeletter}. On the other hand, the lower panel illustrates
a singly peaked profile. The rapid rise and slow decline of the profile
are quite typical for superhumps of SU UMa stars.

\subsection{TSS J0222 as a WZ Sge star}

\begin{figure}
\begin{center}
\resizebox{80mm}{!}{\includegraphics{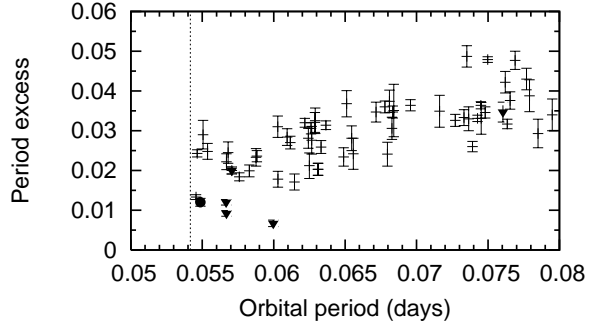}}
\end{center}
\caption{${\epsilon}-$period diagram of SU UMa-type dwarf novae. TSS
 J0222 and confirmed WZ Sge stars are also included with the filled
 circle and filled triangles, respectively. The figure is drawn based on
 table 9 of \citet{pat05suuma}. The vertical dotted line indicates the
 observed period minimum of SU UMa stars on the secular evolution path.}
\end{figure}

The present superoutburst of TSS J0222 showed two types of humps, a
rebrightening, and a large-amplitude of the eruption larger than 6 mag,
all of which are characteristic of WZ Sge stars
(\cite{how95TOAD}; \cite{ima05gocom}). Judging from the
results mentioned before, we can reasonably conclude that TSS J0222 is a
new member of WZ Sge stars with the shortest period ever known. We
further investigated the mass ratio of the object based on an empirical
relation derived by \citet{pat98evolution} as follows:

\begin{equation}
\epsilon = \frac{0.23q}{1+0.27q},
\end{equation}

where ${\epsilon}$ and $q$ are the fractional superhump excess and mass
ratio, respectively. Because the radial velocity study has not been
performed for TSS J0222, the orbital period of TSS J0222 cannot be
exactly specified. However, taking into account the fact
that the period of the early superhump is almost the same as that of the
orbital period (\cite{pat98egcnc}; \cite{pat02wzsge};
\cite{ish03hvvir}), we can validly use the early superhump period above
obtained (0.054868 days) as the orbital period of TSS J0222. Thus the
fractional superhump excess is estimated as being 0.0123(23) with a
little algebra. Substituting this value into the equation (1), we
obtained q = 0.054(10). The value is quite a similar to that listed in
\citet{pat05suuma} for some WZ Sge stars. This implies that
the mass of the secondary star might be as low as 0.075 $M_{\odot}$ even
if the primary white dwarf reaches the Chandrasekhar mass.

Figure 6 shows ${\epsilon}$$-$$P_{\rm orb}$ diagram recently refined by
\citet{pat05suuma}, together with the data of TSS J0222 derived
above. Note that the location of TSS J0222 is close to that of WZ Sge
and AL Com, rather than that of SU UMa-type dwarf novae with short
orbital periods, e.g., WX Cet and SW UMa. This suggests that TSS J0222
bears some resemblance to WZ Sge and AL Com.

\section{Summary}

In this letter, we reported on the newly discovered dwarf
nova, TSS J0222. The obtained light curves provided enough evidence for
early superhumps and genuine superhumps, characteristic of WZ Sge-type
dwarf novae. We derived the superhump period was 0.055544(26) days, as
well as the early superhump period of 0.054868(98) days. The value is
the shortest period among WZ Sge stars. Although a faintness
of the object prevented an accurate calibration of magnitude after the
post-plateau stage, we detected a long rebrightening during this stage,
indicative of the WZ Sge nature of the system. The overall
rebrightenings was classified into ``type-A'' in the
letter, similar to those observed in the 1995 superoutburst of AL Com.
Assuming that the early superhump period is exactly the same as the
orbital period, we roughly derived the mass ratio of TSS J0222 to be
0.054(10). This may indicate the mass of the secondary star is low, like
WZ Sge and AL Com. In the future, further observations with a large
telescope should be performed to accurately determine the orbital period
and the mass ratio of the system.

\vskip 5mm

We would express our gratitude to VSNET observers. This work is
supported by Grants-in-Aid for the 21st Century COE ``Center for
Diversity and Universality in Physics'' from the Ministry of Education,
Culture, Sports, Science and Technology (MEXT), and also by
Grants-in-Aid from MEXT (No. 13640239, 16340057, 16740121,
17740105).

\end{document}